\documentclass[%
 aip,
 amsmath,amssymb,
preprint, 
]{revtex4-1}
\usepackage{geometry} 
\usepackage{graphicx}
\usepackage{epstopdf}
\usepackage{subcaption}
\usepackage{soul}
\usepackage{color}

\begin{document}

\title{Neutron Dark-Field Imaging with Edge Illumination}
\author{Marco Endrizzi}
\email{m.endrizzi@ucl.ac.uk}
\affiliation{Department of Medical Physics and Biomedical Engineering, University College London, Gower Street, London WC1E 6BT, United Kingdom}
\author{Gibril K. Kallon}
\affiliation{Department of Medical Physics and Biomedical Engineering, University College London, Gower Street, London WC1E 6BT, United Kingdom}
\author{Triestino Minniti}
\affiliation{United Kingdom Atomic Energy Authority, Culham Science Centre, Abingdon OX14 3DB, United Kingdom}
\author{Rolf Br\"onnimann}
\affiliation{Empa, Swiss Federal Laboratories for Materials Science and Technology, TNI Lab, \"Uberlandstrasse 129, 8600 D\"ubendorf, Switzerland}
\author{Alessandro Olivo}
\affiliation{Department of Medical Physics and Biomedical Engineering, University College London, Gower Street, London WC1E 6BT, United Kingdom}

\begin{abstract}
We report on an Edge Illumination setup enabling neutron dark-field imaging where two amplitude modulators are used to structure and subsequently analyze the neutron beam. The modulator and analyzer are manufactured by laser ablation of readily available thin metal foils. The sample representation in terms of transmission and dark-field contrast is extracted by numerically inverting a convolution model for the intensity modulation function which had a visibility exceeding 80\%. Two test samples are presented to show how dark-field contrast can complement the more conventional neutron radiography, in particular to investigate the micro-structure of materials. Thanks to the simplicity of the setup, the negligible coherence requirements and the robustness of the method, this approach may find application in multi-contrast neutron radiography and tomography.
\end{abstract}

\maketitle

Neutron imaging techniques produce planar and three-dimensional representations of the interaction of a neutron probe with the sample under investigation. Neutron radiography is arguably the simplest of such approaches, where image contrast arises from differences in attenuation that the neutrons experience whilst traveling through the sample volume \cite{vontobel2006neutron}. Neutron scattering is nowadays an established method for investigating samples at short length scales. It typically entails probing a relatively large volume, of several cubic centimetres, over which the measurements are necessarily averaged \cite{}.
Techniques are evolving rapidly and by means of grating interferometers \cite{pfeiffer2006neutron,sarenac2018three} it is now possible to investigate the sample's transmission and scattering simultaneously and in a spatially resolved fashion. This yields images reporting information spanning across a wide range of length scales, from the centimeter down to the micron and below \cite{grunzweig2008neutron,strobl2008neutron,pushin2017far}. We report here on a neutron dark-field imaging approach which we developed building on our previous work with incoherent X-ray radiation \cite{endrizzi2014hard}. Edge Illumination (EI) is based on the strong structuring of the radiation field by means of an amplitude modulator which is used in conjunction with a harmonically matched amplitude analyzing structure. This setup enables phase sensitivity and the simultaneous measurement of transmission, refraction and scattering properties of a sample. The most common design uses line apertures, in such a way that beam shaping and analysis are one-dimensional, and the same holds for refraction and scattering sensitivity. EI has been extensively studied with X-ray radiation, both at synchrotrons and in laboratory settings, and has shown promise for diverse applications of planar and three-dimensional imaging including mammography, intraoperative imaging, composite materials studies and security screening. With this study we aim at translating the robustness, sensitivity and achromaticty properties of EI to a set-up for application with neutron beams.
\begin{figure*}
\centering
\hfill
\begin{subfigure}[t]{0.6\textwidth}
\includegraphics[width=\linewidth]{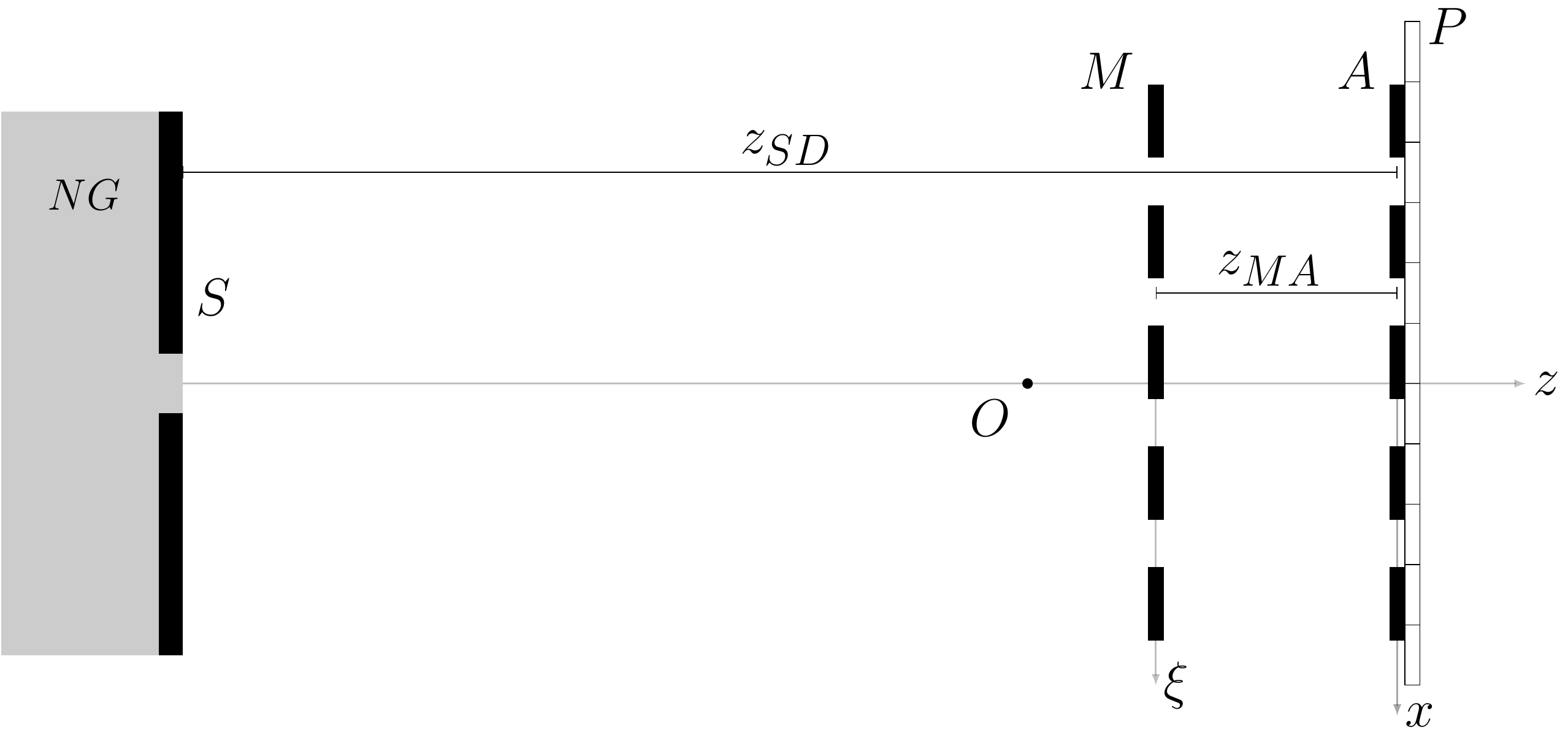}
\caption{\label{fig1a}}
\end{subfigure}
\hfill
\begin{subfigure}[t]{0.33\textwidth}
\includegraphics[width=\linewidth]{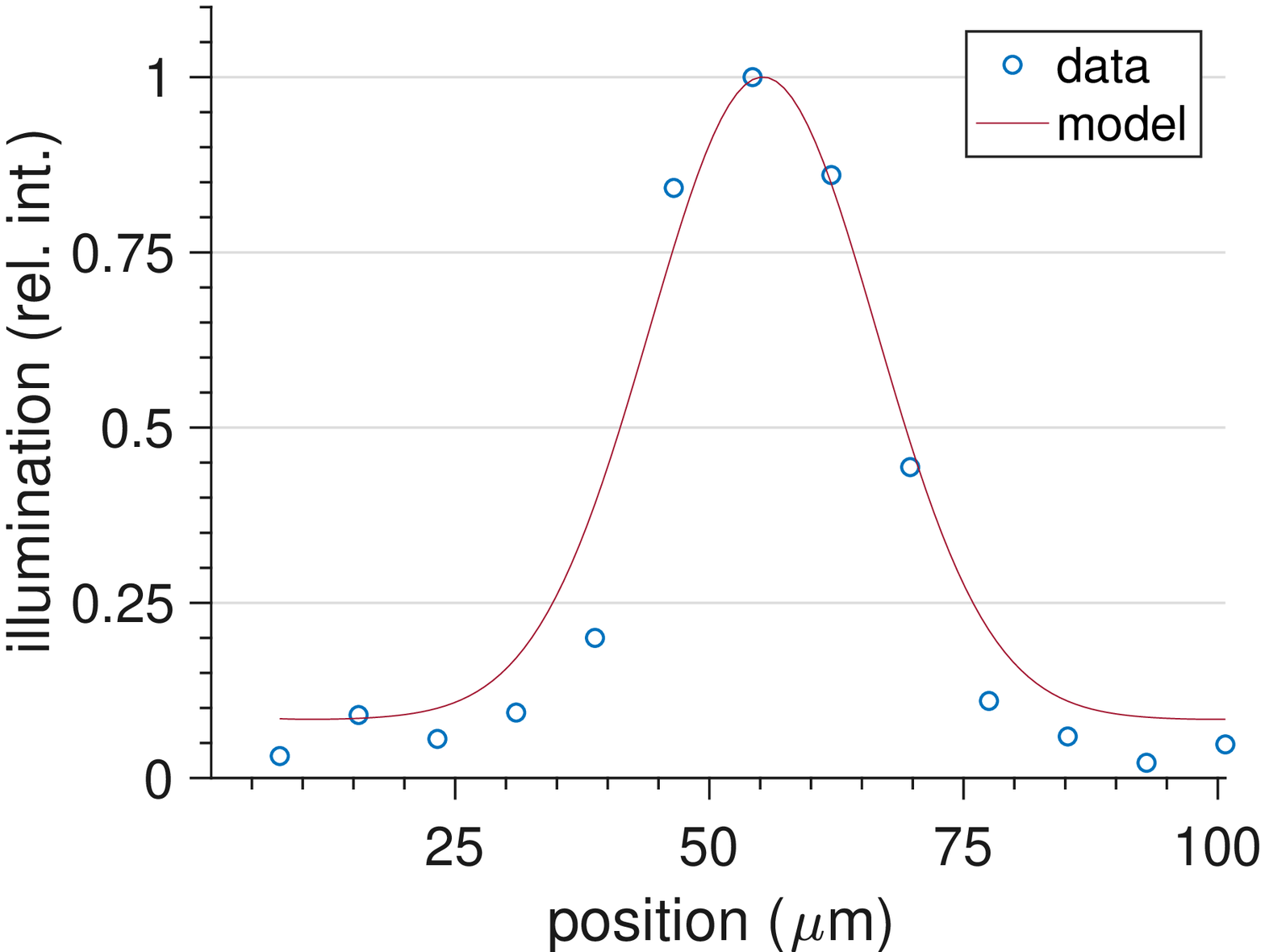}
\caption{\label{fig1b}}
\end{subfigure}
\hfill
\null
\caption{\label{fig:1} (a) Experimental set-up: the beam transported through a neutron guide ($NG$) is collimated by a pinhole ($S$) defining the source size. Amplitude modulator ($M$) and analyzer ($A$) are placed downstream of the imaged object ($O$) and immediately before the detector pixels ($P$). (b) Plot of the experimentally measured and the modeled illumination function.}
\end{figure*}

The experiment was carried out at the neutron imaging and diffraction facility IMAT \cite{minniti2018characterization}, a beamline at the ISIS pulsed neutron spallation source, United Kingdom. 
By exploiting the pulsed structure of the neutron source, IMAT enables both imaging and diffraction applications by making use of time-of-flight techniques for neutron energy discrimination. The neutron spectrum is peaked at $2.5$ \r{A} with a wavelength range extending between $1$ and $7$ \r{A}. The experimental apparatus is depicted in Figure \ref{fig1a}. The beam is transported from the target station 2 through a $46$ m long flight path with a neutron guide and is collimated with a pinhole at the entrance of the experimental hutch. It is then propagated in vacuum and through a series of slits for a further $10$ m, it interacts with the sample and is subsequently modulated and analyzed. Its intensity is finally recorded by the detector which is composed of a 6LiF/ZnS:(Cu/Ag) scintillator optically coupled to a $2048 \times 2048$ pixel Andor Zyla sCMOS 4.2 plus camera. Two test samples were used, the first one was composed of a brass rod $5$ mm in diameter and two layers of foams, Alumina
($Al_2O_3$; 26 pores/cm) and Silicon Carbide  Silicon ($SiC$ 90 \%, $SiO_2$ 7\%, $Al_2O_3$ 1.5\%, $MgO$ 1.5\%; 24 pores/cm). The second one was made of two cylindrical vials, one filled with $D_2O$ and one filled with $D_2O$ and a suspension of borosilicate microspheres, polydispersed in the $9-13$ $\mu$m diameter range.

Image formation in EI can be formulated by means of the Illumination Function (IF), which describes how the intensity recorded by one pixel changes as a function of the relative position of the two apertures. This intensity is maximum when apertures are aligned and reaches a minimum when apertures overlaps absorbing septa. A geometrical optics description models the intensity modulation through the nested convolution between an aperture in the modulator, an aperture in the analyzer and the source intensity distribution demagnified by the factor $z_{MA} / (z_{SD}-z_{MA})$, where $z_{SD}$ is the distance between the pinhole and the detector and $z_{MA}$ is the distance between the modulator and the analyzer. This intensity pattern is usually fitted with a multi-Gaussian distribution and three terms were used in this case. A comparison between the model IF and the experimental data can be seen in Figure \ref{fig1b}. The presence of the sample is taken into account through an additional convolution

\begin{equation}\label{eq:1}
IF_s (\xi) = t \int IF(\xi)O(\xi-\xi')d\xi' 
\end{equation}

where $IF_s$ represents the illumination function in the presence of the sample, $O$ its scattering distribution and $t$ the fraction of intensity it transmits. Equation \ref{eq:1} holds for every image pixel independently.

Modulator and analyzer were fabricated by laser ablation of $50$ $\mu$m thick Gadolinium foils. This thickness was chosen as a good compromise between ease of fabrication, which favors thinner metal foils, and the rigidity of the structure, which favors thicker metal layers to ensure that the resulting structure is stable and self-sustaining. This also enables working with substrate-free optical elements, therefore maximizing transmission through the apertures. The pulse of the ablation laser had a duration of $10$ ps with a maximum energy of 30 $\mu$J at a wavelength of $355$ nm. The pulses were directed via a scan head towards the sample which was mounted on xy-stage. Modulator and analyzer were produced by combination of scan head operation and stitching to achieve high large-scale precision. The fabrication process partially affected the flatness of the foils which were slightly warped as a result. We note that this is not a problem for EI imaging. Since we are using a complete IF scan, the modulator and the analyzer do not require a high degree of alignment accuracy because each pixel is treated independently and misalignment does not affect the local IF visibility \cite{endrizzi2015laboratory}. 
The system design features $p_M = 109$, $p_A = 109.6$ $\mu$m pitches and $a_M = 12$, $a_A = 13.3$ $\mu$m apertures for the modulator ($_M$) and analyzer ($_A$) respectively. We defined the source size with a pinhole of $20$ mm diameter, and the system geometry was such that $z_{MA} = 31$ mm and $z_{SD} = 10$ m.

We achieved the modulation by shifting $M$ along the $\xi$ axis (Fig. \ref{fig1a}) in steps by means of a piezoelectric linear stage. The experimentally measured IF is shown in Figure \ref{fig1b} and its visibility
\begin{equation}
V = \frac{IF_{max} - IF_{min}}{IF_{max}+IF_{min}}
\end{equation}
is $82\%$. The Gadolinium thickness is sufficient to safely assume a $100 \%$ absorption efficiency at all wavelengths. Combined with the absence of a substrate, this ensures that the system is completely achromatic and that the entire radiation spectrum is simultaneously exploited for imaging. The only energy-dependent modulation left in the system is the one introduced by the sample. $IF$ and $IF_s$ scans were acquired by taking $13$ exposures, $180$ s each, with a shift of $\Delta \xi = 7.75$ $\mu$m between modulator and analyzer at each step. Transmission ($t$) and dark-field images (variance of $O$ distribution) are then extracted by measuring, on a pixel-by-pixel basis, the reduction of the total area under the IF and its broadening, respectively. This is achieved through a local non-linear fit procedure that assumes a Gaussian form for the sample's scattering distribution $O$.
\begin{figure}[]
\centering
\begin{subfigure}[t]{0.3\textwidth}
\includegraphics[width=\linewidth]{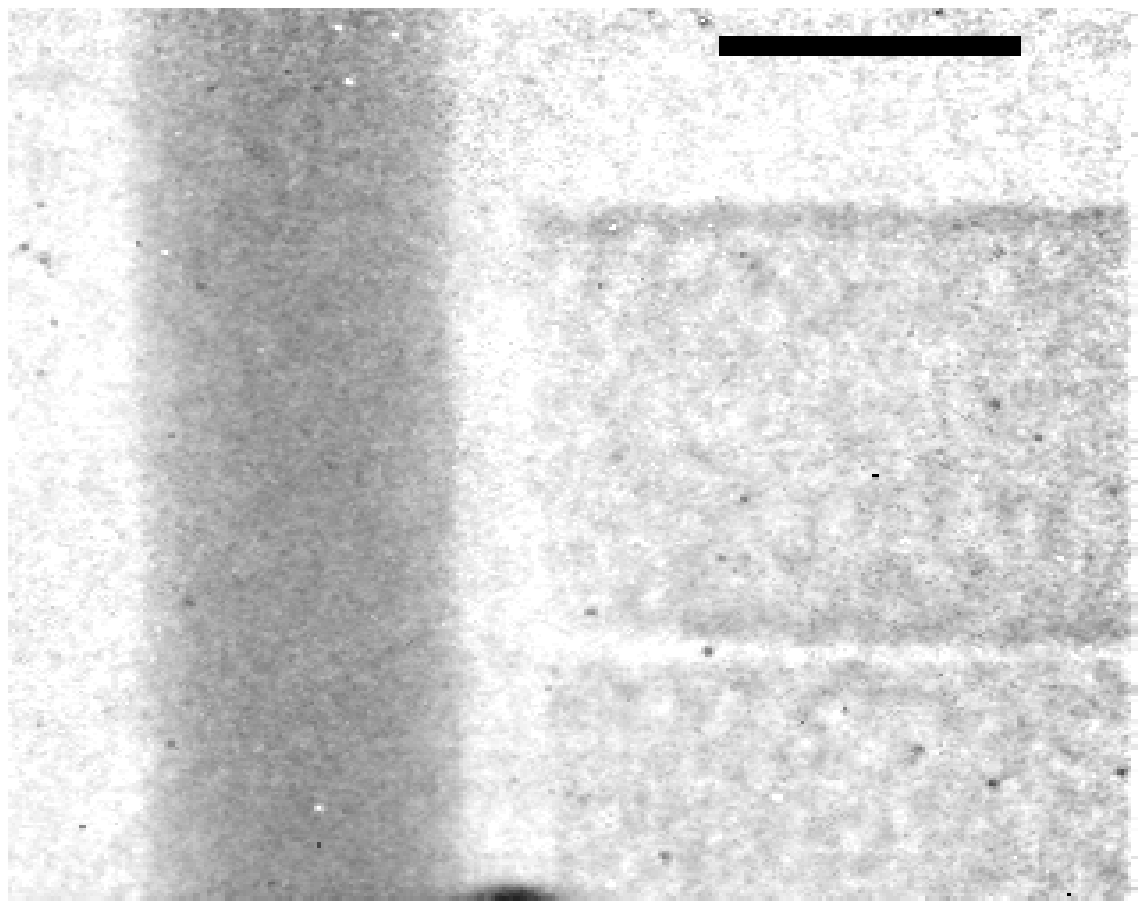}
\caption{\label{fig2a}}
\end{subfigure}
\begin{subfigure}[t]{0.3\textwidth}
\includegraphics[width=\linewidth]{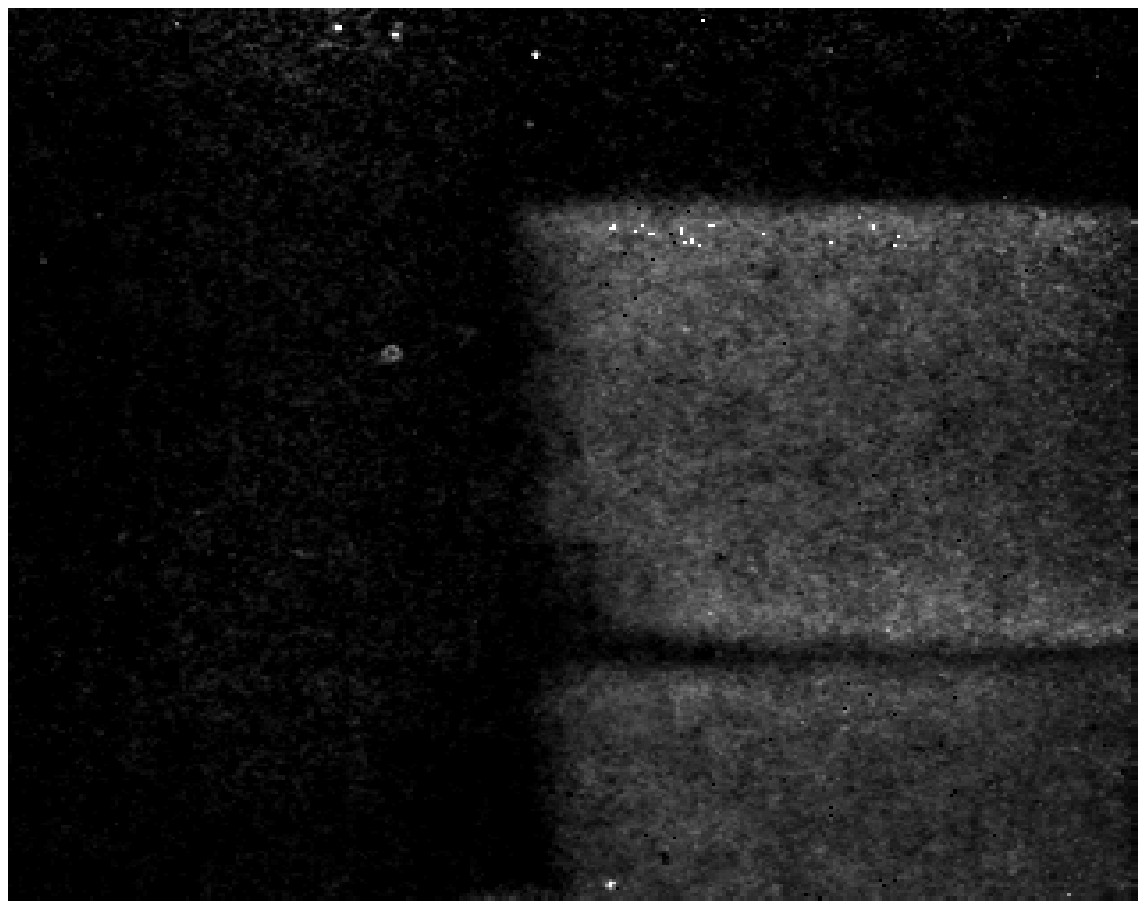}
\caption{\label{fig2b}}
\end{subfigure}
\caption{\label{fig:2}First test sample: (a) transmission and (b) dark-field images. The brass rod (left-hand side) and the alumina/Silicon carbide foams (right-hand side, below and above, respectively) are both visible in transmission contrast while only the foams appear bright in the dark-field contrast. The scale bar in the top-right corner of panel (a) represents $5$ mm.}
\end{figure}

Results from the first test sample are shown in Fig.\ref{fig:2}, with transmission and dark-field images in panels (\ref{fig2a}) and (\ref{fig2b}), respectively. The brass rod and both foams (alumina and Silicon carbide) are visible in the transmission image whilst only the foams can be seen in the dark-field image. This provides a qualitative insight on the micro-structure discrimination capabilities enabled by the concurrent availability of these two contrast channels. 
\begin{figure*}
\centering
\hfill
\begin{subfigure}[t]{0.3\textwidth}
\includegraphics[width=\linewidth]{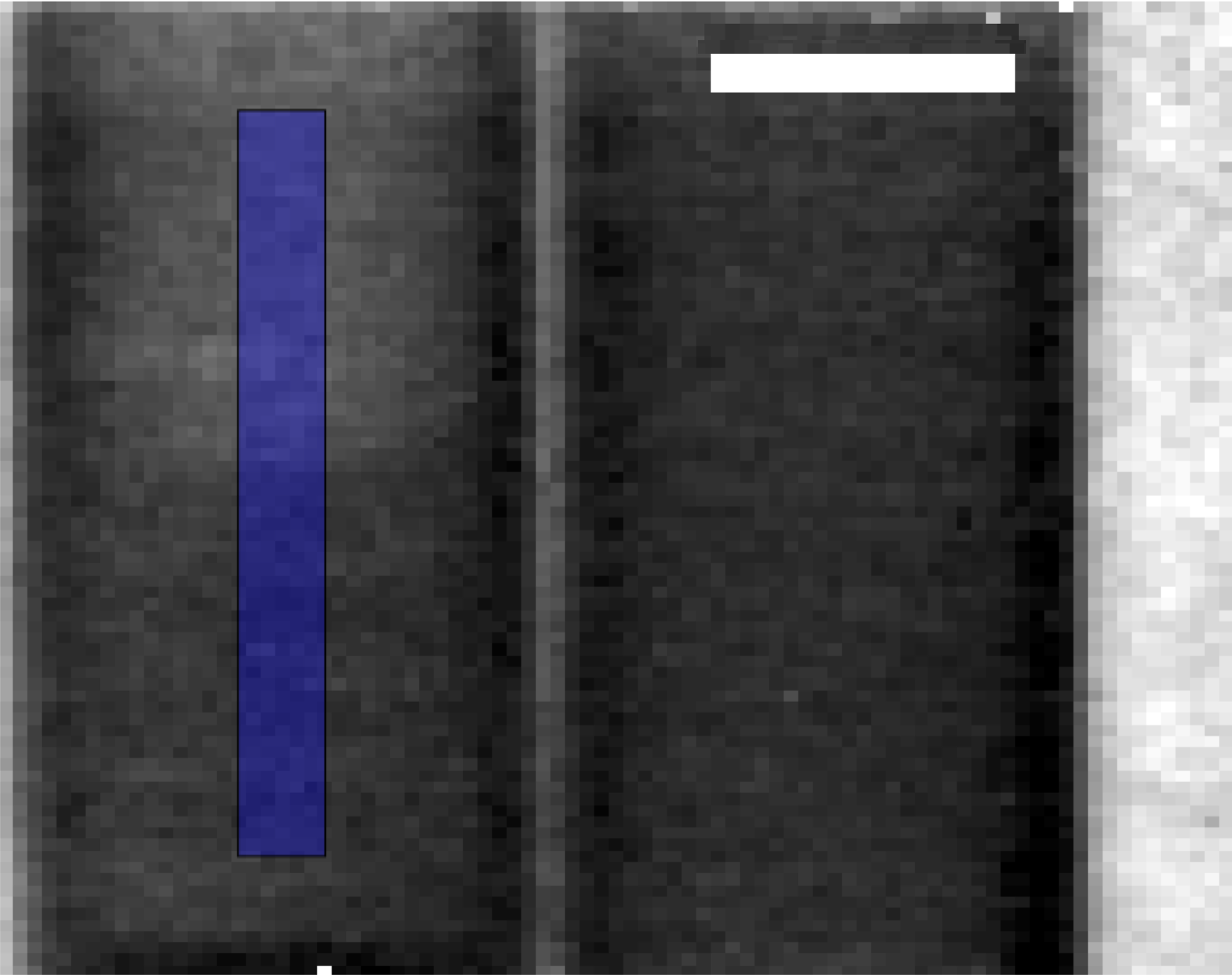}
\caption{\label{fig3a}}
\end{subfigure}
\hfill
\begin{subfigure}[t]{0.3\textwidth}
\includegraphics[width=\linewidth]{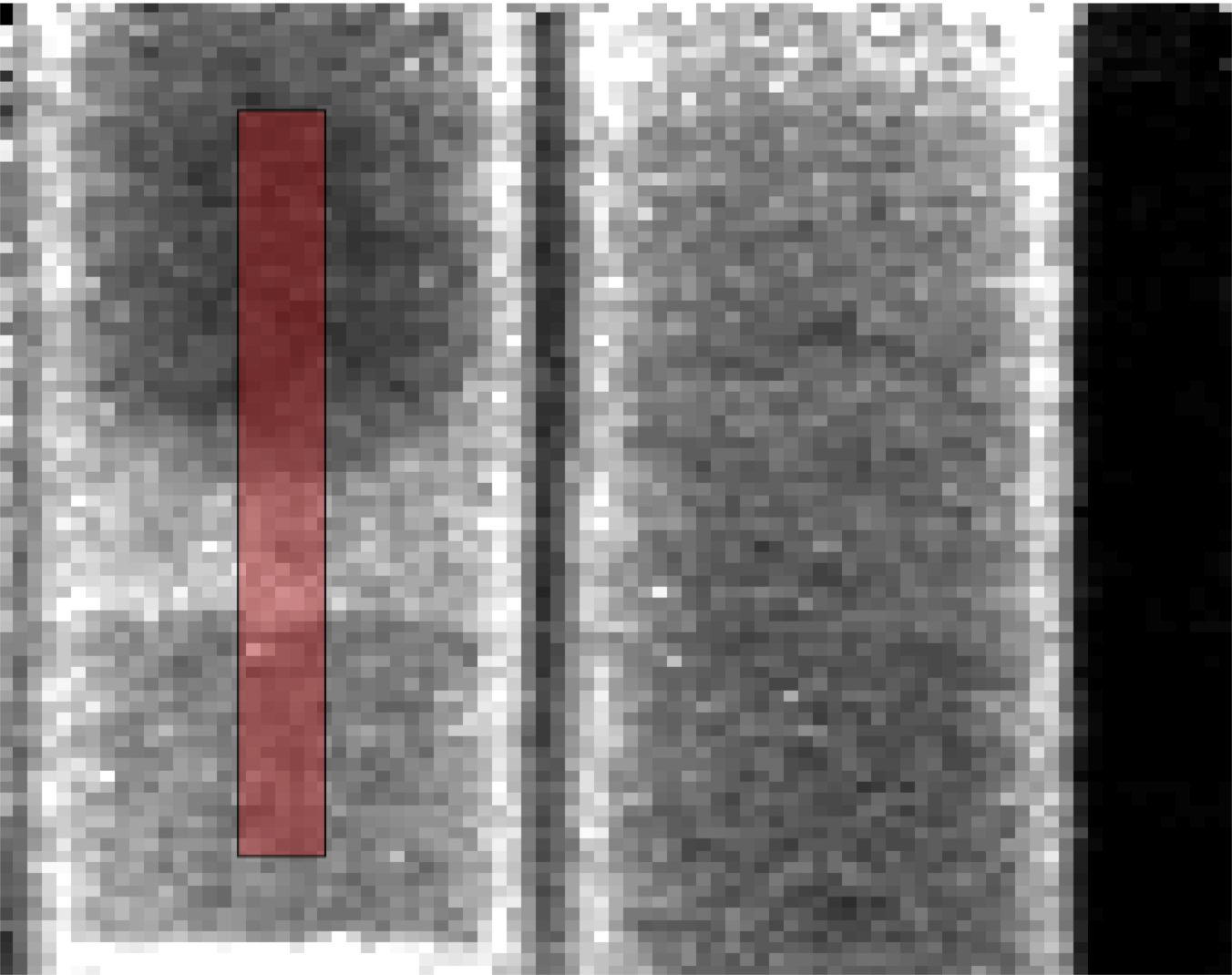}
\caption{\label{fig3b}}
\end{subfigure}
\hfill
\begin{subfigure}[t]{0.33\textwidth}
\includegraphics[width=\linewidth]{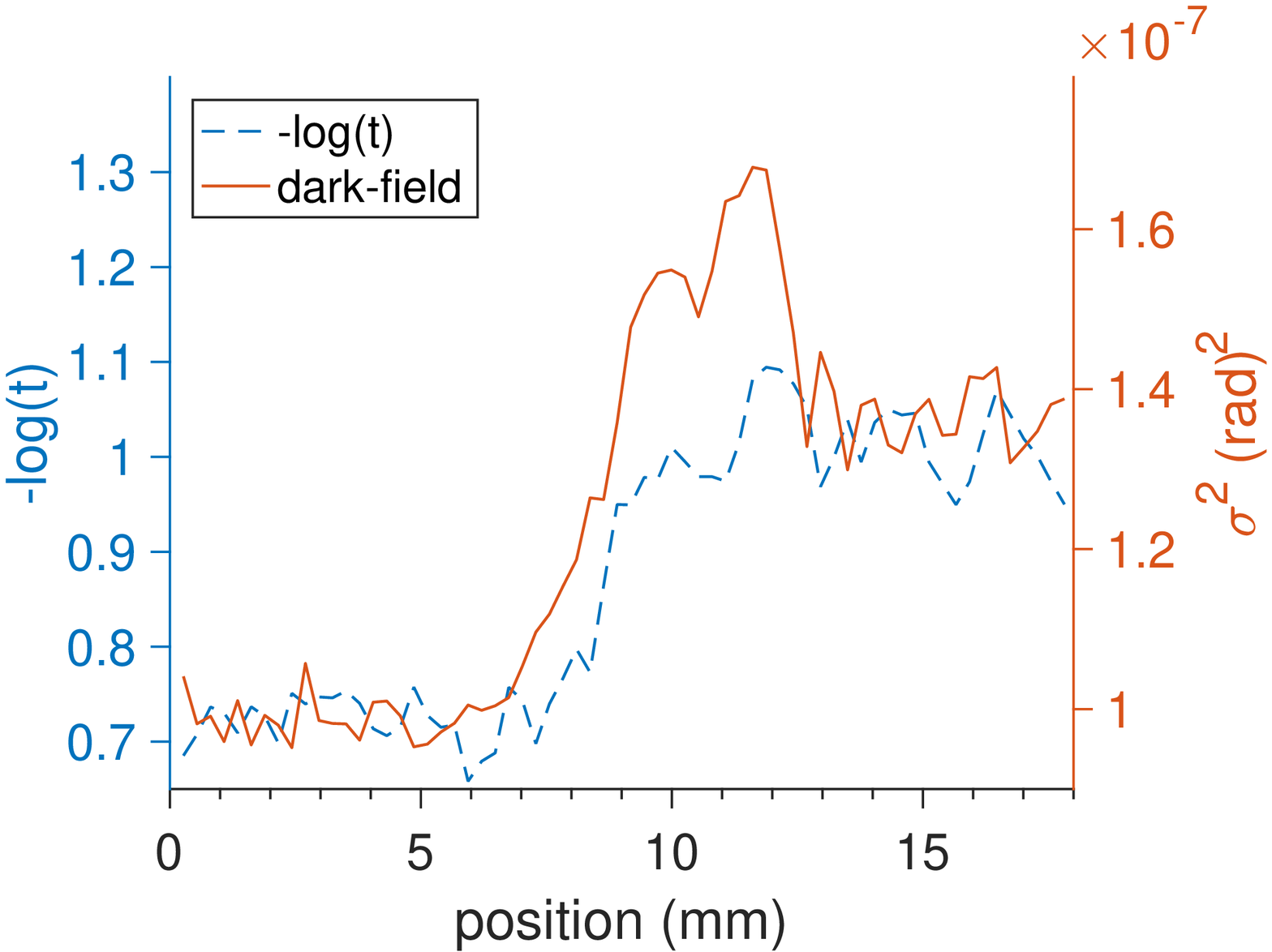}
\caption{\label{fig3c}}
\end{subfigure}
\hfill
\null
\caption{\label{fig:3} Second test sample: (a) transmission and (b) dark-field images. The vial containing the micro-spheres suspension exhibits a transition region near the liquid surface which is clearly visible in the dark-field image. Panel (c) shows a comparison of the line profiles measured along the regions highlighted in panels (a) and (b) for a quantitative assessment. The scale bar in the top-right corner of panel (a) represents $5$ mm.}
\end{figure*}
Images of the second test sample are shown in Fig. \ref{fig:3}. Transmission images of the microspheres suspension (left) and $D_2O$ only (right) can be seen in panel (\ref{fig3a}). Similarly, dark-field images can be visually compared by looking at  panel (\ref{fig3b}). Panel (\ref{fig3c}) shows the linear plots measured in the regions highlighted in blue and red, for the transmission and dark-field image, respectively. The interface between liquid and air is well visible in both the transmission (blue dashed line) and the dark-field (red continuous line) images. However, a transition region is clearly identifiable by looking at the dark-field image, whilst it is much less, if at all, visible in the transmission contrast channel. We interpret this signal as an indication of the presence of a layer, in the vicinity of the liquid surface, where the packing density of spheres changes rapidly.

Some bright and dark spots are visible in the images in Fig. \ref{fig:2} and Fig. \ref{fig:3}, in a way that resembles salt and pepper noise. This is due to the imperfect stability of the retrieval algorithm that does not converge in some pixels and returns largely incorrect values. These localized problems are linked to the presence of a relatively high noise in the raw images which was due to the need to run the camera without cooling, for a temporary problem with its vacuum  that occurred during the experiment.

In summary, we reported on an experimental setup for neutron dark-field imaging based on Edge Illumination. The experimental apparatus makes use of an amplitude modulator and an amplitude analyzer, offers high ($>80\%$) intensity modulation visibility and is completely achromatic. We showed, through experimental results on two test samples, examples of how dark-field contrast can complement standard attenuation contrast, enabling the simultaneous extraction of additional information that can be useful in characterizing the micro-structure of materials. We found these first proof-of-principle results encouraging towards the development of a fully achromatic neutron dark-field imaging system that can efficiently operate across a wide wavelength spectrum.

\section*{Acknowledgements} 
ME was supported by the Royal Academy of Engineering under the RAEng Research Fellowships scheme. AO was supported by the Royal Academy of Engineering under the Chairs in Emerging Technologies scheme. We thank ISIS Neutron and Muon source for access to the IMAT beamline with proposal 1900012 that was integral to the results presented here.

\section*{Data Availability Statement} 
The data that support the findings of this study are available from the corresponding author upon reasonable request.

\bibliography{./references}

\end{document}